\documentclass[conference]{IEEEtran}
\IEEEoverridecommandlockouts
\usepackage{cite}
\usepackage{amsmath,amssymb,amsfonts}
\usepackage{algorithmic}
\usepackage{graphicx}
\usepackage{textcomp}
\usepackage{xcolor}
\def\BibTeX{{\rm B\kern-.05em{\sc i\kern-.025em b}\kern-.08em
    T\kern-.1667em\lower.7ex\hbox{E}\kern-.125emX}}
    
\usepackage[binary-units]{siunitx}
\usepackage{stfloats}
\usepackage{cuted}
\begin{document}
\thispagestyle{empty}
\newpage
\onecolumn
\begin{center}
This paper has been accepted for publication in 2020 IEEE OES OCEANS.
\vspace{0.75cm}\\
DOI: \\ % \href{https://doi.org/10.1109/ICRA.2019.8793718}{\textcolor{blue}{10.1109/ICRA.2019.8793718}}\\
IEEE Xplore: \\% \href{https://ieeexplore.ieee.org/document/8793718}{\textcolor{blue}{https://ieeexplore.ieee.org/document/8793718}}
\vspace{1.25cm}
\end{center}
©2020 the authors under a Creative Commons Licence CC-BY-NC-ND. Personal use of this material is permitted. Permission from IEEE must be obtained for all other uses, in any current or future media, including reprinting/republishing this material for advertising or promotional purposes, creating new collective works, for resale or redistribution to servers or lists, or reuse of any copyrighted component of this work in other works.
\twocolumn

\title{Single Image Super-Resolution for Domain-Specific Ultra-Low Bandwidth Image Transmission
}

\author{\IEEEauthorblockN{Jesper Haahr Christensen}
\IEEEauthorblockA{\textit{DTU Electrical Engineering} \\
\textit{Technical University of Denmark}\\
\textit{AUV Competence Centre} \\
\textit{ATLAS MARIDAN ApS} \\
{\tt \small jehchr@elektro.dtu.dk}}
\and
\IEEEauthorblockN{Lars Valdemar Mogensen}
\IEEEauthorblockA{\textit{AUV Competence Centre} \\
\textit{ATLAS MARIDAN ApS}\\
2960 Rungsted Kyst, Denmark \\
{\tt \small lvm@atlasmaridan.com}}
\and
\IEEEauthorblockN{Ole Ravn}
\IEEEauthorblockA{\textit{DTU Electrical Engineering} \\
\textit{Technical University of Denmark}\\
2800 Kgs. Lyngby, Denmark \\
{\tt \small or@elektro.dtu.dk}}
}

\maketitle

\begin{figure*}[b!]
    \centering
    \includegraphics[width=1\linewidth]{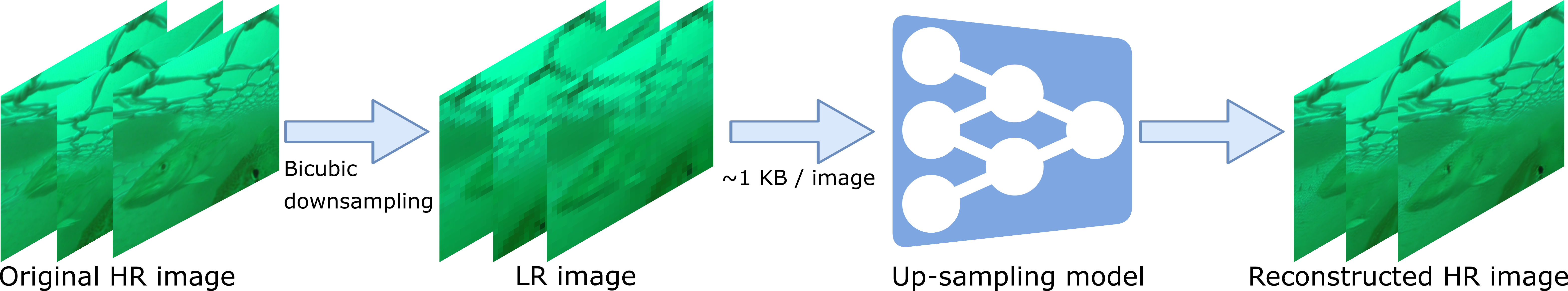}
    \caption{\textit{Method overview.} The original HR image is bicubic down-sampled to a small LR image of about \SI{1}{\kilo\byte} in size. The LR image -- small enough to be transmitted over low-bandwidth communication channels -- is input to a neural network trained to reconstruct the HR image. The output image shows visually pleasing results and even recovers high-frequency details.}
    \label{fig:overview}
\end{figure*}

\begin{abstract}
Low-bandwidth communication, such as underwater acoustic communication, is limited by best-case data rates of 30--50 kbit/s. This renders such channels unusable or inefficient at best for single image, video, or other bandwidth-demanding sensor-data transmission. To combat data-transmission bottlenecks, we consider practical use-cases within the maritime domain and investigate the prospect of Single Image Super-Resolution methodologies. This is investigated on a large, diverse dataset obtained during years of trawl fishing where cameras have been placed in the fishing nets. We propose down-sampling images to a low-resolution low-size version of about 1 kB that satisfies underwater acoustic bandwidth requirements for even several frames per second. A neural network is then trained to perform up-sampling, trying to reconstruct the original image. We aim to investigate the quality of reconstructed images and prospects for such methods in practical use-cases in general. Our focus in this work is solely on learning to reconstruct the high-resolution images on ``real-world'' data. We show that our method achieves better perceptual quality and superior reconstruction than generic bicubic up-sampling and motivates further work in this area for underwater applications. 
\end{abstract}


\section{Introduction}
Single Image Super-Resolution (SISR) aims at recovering a high-resolution (HR) image from a single low-resolution (LR) one. That is to reconstruct a HR image with high Peak Signal-to-Noise Ratio (PSNR) and comparable perceptual pleasantness to the original HR image. Various methods to accomplish this already exist and are continuously evolved. 
An early solution that remains generally applied is using interpolation. This is to increase the spatial size of a LR image and estimate missing pixel values by some interpolation method, e.g. nearest neighbor, bilinear, or bicubic. However, while being a popular choice, these interpolation methods lack the ability to generate high-quality results. The images tend to become blurry and are without sufficient high-frequency details. 
Deep neural networks have recently become popular for SISR and have seen continuous improvements over the past years. Various convolutional neural network (CNN) architectures and loss-function designs have improved SISR performance drastically \cite{10.1007/978-3-319-10593-2_13,inproceedings,article_harris,8099502,inproceedings_residual,10.1007/978-3-030-01234-2_18,10.1007/978-3-030-11021-5_5}. Current state-of-the-art points towards using residual learning \cite{resnet} and deep Generative Adversarial Networks (GANs)~\cite{Goodfellow:2014:GAN:2969033.2969125} to obtain a high PSNR while not producing over-smoothed results and thus recover a visually pleasant HR image with great detail of high-frequency components.  

Building on recent work within SISR, we propose using such methods for practical use-cases in the maritime domain. We seek to obtain a model that will allow for ultra-low bandwidth data transmission. The specific use-case in mind for this work is domain-specific image transmission using underwater acoustic links during trawl fishing operations. We investigate the theoretical prospect w.r.t. bandwidth utilization and HR image reconstruction quality. An overview of our method is shown in Fig.~\ref{fig:overview}. Here an original HR image is bicubic down-sampled to a LR image thumbnail with a size footprint of about \SI{1}{\kilo\byte}. This is within bandwidth limitations of most modern high-speed mid-range acoustic modems for one or several image transfers per second. This \SI{1}{\kilo\byte} LR image is input to our neural network, which predicts a HR image with spatial image-resolution scale-factors up to $\times 8$ compared to the LR image.

The dataset used for training our model for trawl fisheries has been supplied by the German Thünen Institute of Baltic Sea Fisheries. This consists of many years of video data that has been obtained during trawl-specific research operations. % on the institute's research vessel Clupea. 

In contrast to previous approaches found in literature \cite{6747842,8604666}, we do not investigate data modulation schemes or region of interest based compression or modeling techniques. Instead, we learn a mapping from a LR small-sized RGB image to a HR image that compares to the original HR image w.r.t. PSNR and perceptual quality. Our method allows for the full image to be transferred, with no modifications to existing communication protocols and hardware. We provide our model -- building on state-of-the-art SISR -- and show how such methods may be utilized for practical use-cases within the maritime domain for domain-specific low-bandwidth data transmission. We also note that the application domain is not limited to trawl fisheries, and not even to optical imagery. As such, these techniques may be well suited for sonar images or other data formats with high bandwidth requirements.

\section{Related Work}

\noindent\textbf{Single Image Super-Resolution} \quad 
In line with CNNs revolutionizing a wide range of computer vision tasks, CNNs have also found their place in SISR. The first pioneering work is introduced in \cite{7115171}, and uses bicubic interpolation to up-sample a LR image by some scale-factor and then refine the output by a set of convolutional filter kernels. Inspired by this, \cite{8099502} replaced the interpolation method by sub-pixel convolutions \cite{7780576} and turned to residual learning \cite{resnet}. Later, a variety of architectures and loss functions have been introduced to overcome issues related to using only a pixel loss for recovering HR images. Pixel loss tends to produce over-smoothed results and lack detail of high-frequency content. In \cite{inproceedings_perceptual}, a popular perceptual loss is introduced to combat shortcomings of pixel loss. In more recent work, GANs are used to combine the compelling results of using an adversarial network in addition to previously mentioned work \cite{8099502,10.1007/978-3-030-11021-5_5}. Most recent results point towards the use of GANs and residual learning, and as such, we base our architecture on this. \\

\noindent\textbf{Acoustic Image/Video Transmission} \quad 
Earlier work investigates the use of efficient compression algorithms and high-level data modulation \cite{1283454,6747842} with requirements to specific hardware and customized communication protocols. In contrast, we seek to obtain an applicable method that utilizes off-the-shelf commercial underwater communication systems. More recent work \cite{8604666} seeks to segment sonar images into a foreground (containing high-information regions) and a background (containing low-information regions) to highly compress less informative parts of an image, and to maintain high detail of more informative image regions. In \cite{7999241}, the authors propose a learning scheme where a light-weight CNN is used to down-sample the original image to create a more efficient and compact representation over bicubic down-sampling. On the reconstruction-side, the compact image is first up-sampled by bicubic interpolation and then refined by another light-weight CNN similar to \cite{7115171}. In our initial work, we do not focus our attention on compression schemes. We focus on the applicability and reconstruction quality for specific practical use-cases, as our LR image-representation already satisfies constraints put by current underwater communication channels.

\section{Optical Trawl Dataset}
\begin{figure}
    \centering
    \includegraphics[width=1.0\linewidth]{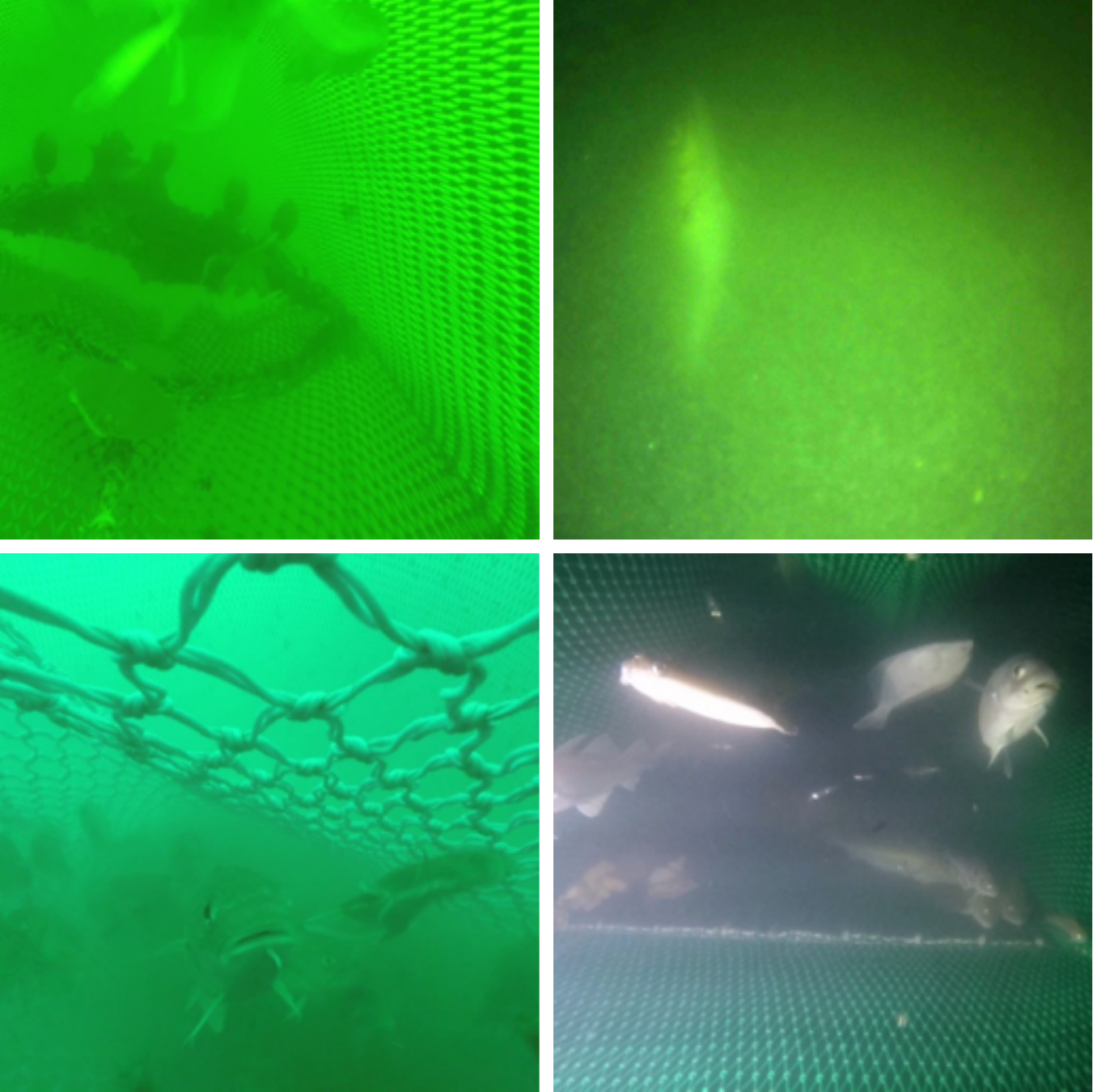}
    \caption{\textit{Samples from the dataset.} The large dataset contains diverse images with varying quality. As exemplified here, some images lack color, visibility or are affected by mud from the seabed. Some images also lack sufficient lighting which in turn leads to longer exposure times, and hence blurry images.}
    \label{fig:dataset}
\end{figure}

The dataset used in this work has been kindly supplied by the Thünen Institute of Baltic Sea Fisheries. It comprises compressed video recordings obtained during years of trawling -- a fishing method where a boat deploys a large net and drags it after the boat -- ranging back from 2014 to 2019. Using the research vessel of Thünen, Clupea, GoPro cameras have been attached to the fishing net for catch inspections. The position of the camera(s) changes for each trawl; hence, the material contains large diversities over many different viewing angles from the net entrance to the cod end. Trawling is carried out for many different fish species for pelagic fisheries (water column trawl) and demersal fisheries (bottom trawl). The quality of the videos is varying due to the effective camera sensors, settings, and illumination. Some footage contains minimal color and light; others suffer from motion blur, organic detritus (marine snow), or have very low visibility due to material from the seabed (e.g. mud) is being stirred up by the impact of the trawl. In Fig.~\ref{fig:dataset}, we show four samples from our dataset to exemplify this. 

From the large video dataset, we sample about \SI{150000}{} images to obtain a broad representation of different scenarios, settings, fish species, lighting conditions, camera placements, and image quality. The train/test split is separated by trawls to not mix up images in train/test from the same trawl operation.

\section{Proposed Method}

\begin{figure}
    \centering
    \includegraphics[width=1.0\linewidth]{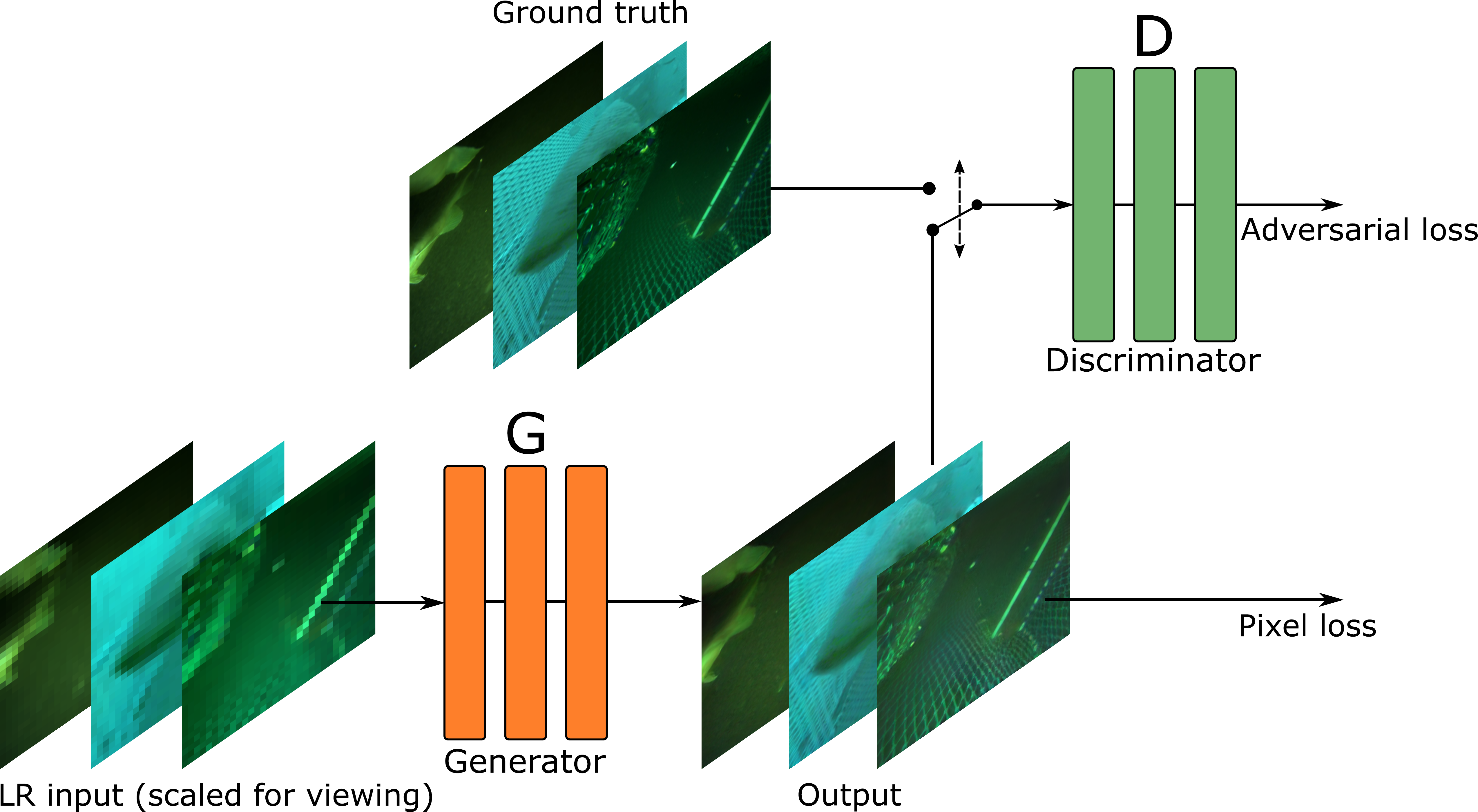}
    \caption{\textit{Overview of network architecture.} The generator $G$ seeks to recover a HR image of the LR input one. The discriminator $D$ compares the generated HR image with the ground truth HR image and enforce better recovery of texture details and more visually pleasing results. 
    \label{fig:model}}
\end{figure}

In order to train a neural network for up-sampling an input from LR image-space to HR image-space, our method defines first a generative model to produce estimations of the ground-truth HR image from the LR input image. Second, we add an adversarial model adding to the loss of our training scheme and guide the generative model to produce more realistic looking and detailed images. 
An overview of our network architecture is shown in Fig.~\ref{fig:model}. 

As we show in Fig.~\ref{fig:model}, the LR input images are processed by a generator, $G$, that aims to recover the HR image. The adversarial discriminator, $D$, is given both the generated HR images and the ground-truth HR images to evolve and encourage the generator to produce more realistic looking images with a perceptual quality superior to those obtained by minimizing a pixel-wise error measurement on the generator alone. Details of architecture and loss functions are given in the following.

\subsection{Generator}
The generator aims to estimate a super-resolved HR image from the LR input image. To do so, we define a CNN architecture building on the generator proposed in ESRGAN~\cite{10.1007/978-3-030-11021-5_5}. We take advantage of residual learning in the LR image-domain, then up-sample our image by an up-sampling function with learnable parameters, and finally refine the image by two sets of convolutions in the HR image-domain. Specifically, we employ a set of 16 Residual-in-Residual Dense Blocks instead of 23 used in~\cite{10.1007/978-3-030-11021-5_5}. We further replace the original interpolation functions with sub-pixel convolutions \cite{7780576} to up-sample from LR to HR and use $\times 8$ scale factor instead of $\times 4$ scale factor as used in ESRGAN. In summary, this results in a ``slimmer'' version of the generator in ESRGAN with the capability to learn the up-sampling functions rather than rely on interpolation, and with increased scaling factor to generate higher-resolution images.

The output of the generator is compared against the ground truth HR images by calculating the pixel-wise L1 error:
\begin{align} \label{eq:lg}
    \mathcal{L}_{L_1}(G) = \mathbb{E}_{x,y}\left[||y-G(x)||_1\right]
\end{align}
where $x$ is the input LR image, $y$ is the HR ground truth image, and $G$ is the generator.

\subsection{Discriminator}
It is well known that optimizing the pixel-wise L1 (or L2) loss on the generator alone produces blurry results and lacks high-frequency detail. To encourage the generator to produce perceptually superior solutions and hence more detailed and realistic images, we add an adversarial discriminator, $D$. We design our discriminator as a Relativistic Average~\cite{jolicoeur2018relativistic} PatchGAN~\cite{8100115}. This motivates the discriminator to model high-frequency structure and therefore only penalize structure at the scale of local image patches. Hence, the discriminator aims to measure if each $N\times N$ image patch is more realistic than the real or fake one. In contrast to the original PatchGAN architecture, we remove all Batch Normalization~\cite{10.5555/3045118.3045167} layers, as these tend to introduce unwanted artifacts in image generation tasks. Further, we introduce relativism in the loss scheme of the discriminator. 
We condition the discriminator on the relativistic patch-wise least-squares error and express the GAN loss as:

\begin{align}
\begin{split} \label{eq:lgd}
    \mathcal{L}_{GAN}(D) = &\mathbb{E}_{y}\left[\|D(y)-\Bar{D}(G(x))-1\|^2_2\right] \\
    +&\mathbb{E}_{x}\left[\|D(G(x))-\Bar{D}(y)+1\|^2_2\right] 
\end{split} \\
\begin{split} \label{eq:lgg}
    \mathcal{L}_{GAN}(G) = &\mathbb{E}_{y}\left[\|D(y)-\Bar{D}(G(x))+1\|^2_2\right] \\
    +&\mathbb{E}_{x}\left[\|D(G(x))-\Bar{D}(y)-1\|^2_2\right] 
\end{split}
\end{align}
where $\Bar{D}(\cdot)$ denotes the average of the discriminator output.

\subsection{Full Objective}
Combining the loss from (\ref{eq:lg}), (\ref{eq:lgd}) and (\ref{eq:lgg}), we can express our full objective as:
\begin{align}
     \min_G \max_D\quad \frac{1}{2} \left[ \mathcal{L}_{GAN}(D)+\mathcal{L}_{GAN}(G)\right]\lambda + \mathcal{L}_{L_1}(G)
\end{align}
where $\lambda$ is a weighting factor.

\section{Experiments}

\subsection{Training Details}
All networks are implemented in PyTorch and trained using $\times 4$ NVIDIA RTX 2080 Ti GPUs. HR images have the spatial size $256\times 256$, and LR images obtain the spatial size $32\times 32$ by down-sampling HR images using bicubic interpolation. LR images are further JPEG encoded, reducing the effective size-footprint from about \SI{3}{\kilo\byte} to about \SI{1}{\kilo\byte}.
As in \cite{10.1007/978-3-030-11021-5_5}, we divide our training process into two stages. First, we train our generator only using pixel-wise L1 loss and thus maximize the PSNR. We then employ this model as initialization for our generator when training our GAN-based model. This is to obtain a more stable training process, as the discriminator will already receive relatively good images from the generator, and not start from pure noise. 
For all models, we use a learning rate for of $1\times 10^{-4}$ and a mini-batch size of 8 per GPU. The Adam~\cite{article} solver is used for optimization with parameters $\beta_1$ and $\beta_2$ set to $0.9$ and $0.999$ respectively. The GAN-based model use the weight parameter $\lambda=1\times 10^{-2}$. 

\subsection{Results}

By pre-training our generator, we obtain an average PSNR of $25.6$, while bicubic up-sampling yields a PSNR of $23.3$. However, as previously described, minimizing only the pixel-wise error yields over-smoothed images that lack high-frequency detail. Hence we train the GAN-based model initialized with the pre-trained generator. The GAN-based model obtains a final PSNR of $24.5$, which is less than the pre-trained generator. This is expected due to the GAN-based model also minimizes the adversarial (perceptual) loss. 

\begin{figure}
    \centering
    \includegraphics[width=1.0\linewidth]{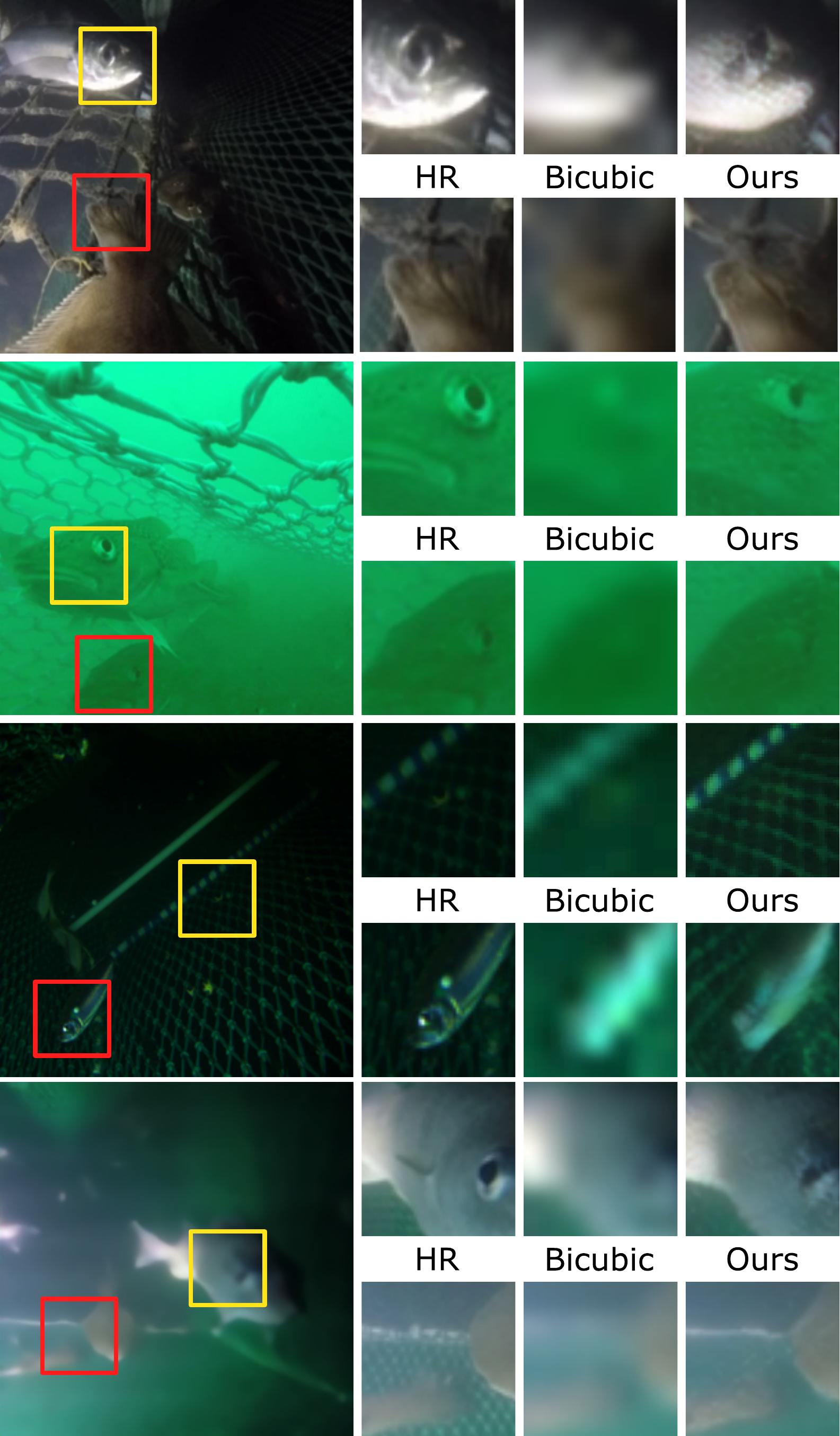}
    \caption{\textit{Qualitative results.} Local image patches are inspected more carefully by enlarging two chosen regions in the images. For each image, the upper row corresponds to the yellow square, and the lower row corresponds to the red square. HR corresponds to the original image patch, bicubic corresponds to a $\times 8$ up-sampling of the LR image using bicubic interpolation. The rightmost column is the result of $\times 8$ up-sampling using our method
    \label{fig:results2}}
\end{figure}

\begin{figure*}
    \centering
    \includegraphics[width=1.0\linewidth]{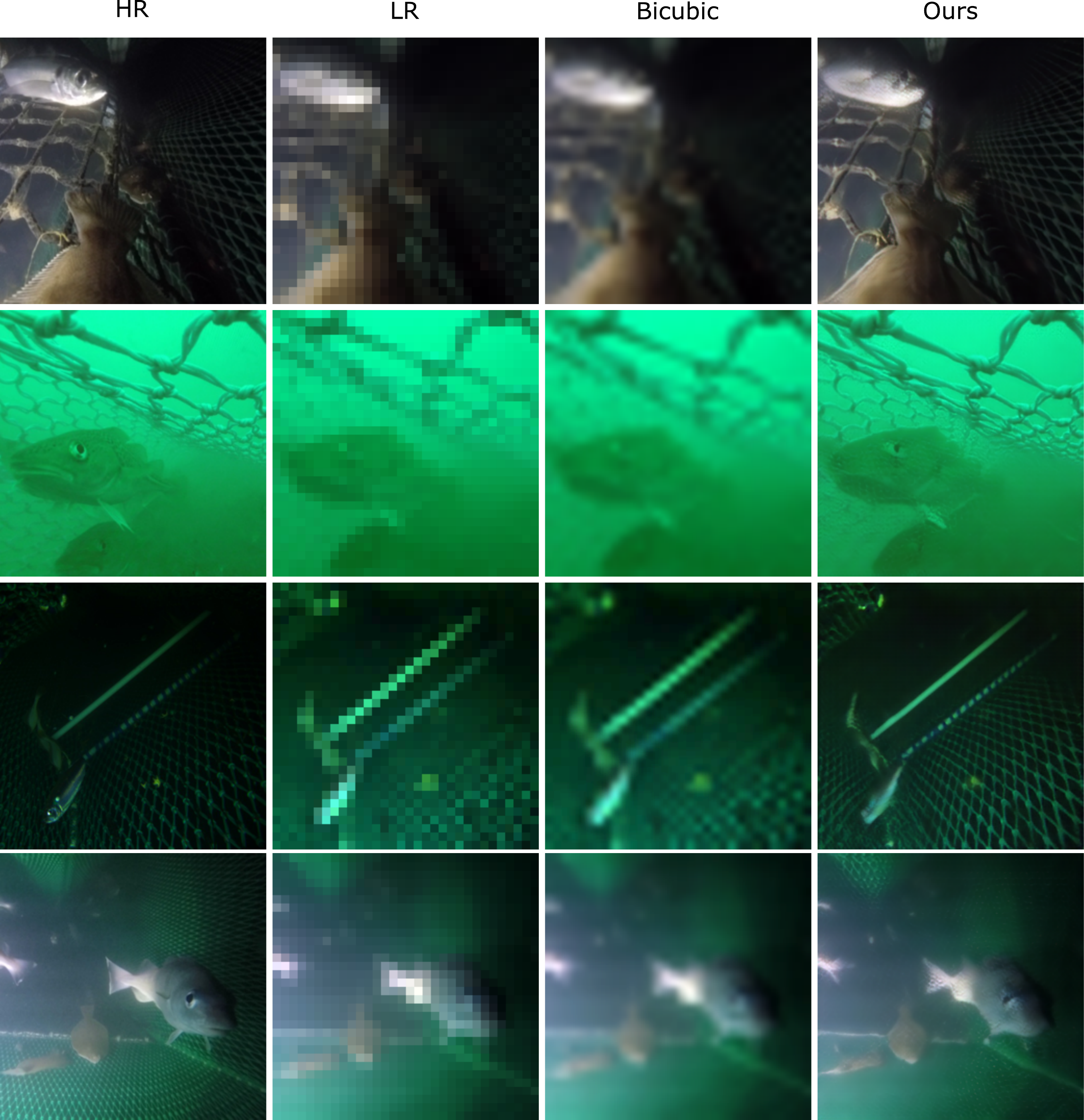}
    \caption{\textit{Qualitative results.} The HR image is a $256\times 256$ bicubic down-sampled version of the original frame extracted from compressed video. We obtain the LR image of $32\times 32$ by bicubic down-sampling of the HR image. The ``Bicubic'' column corresponds to a $\times 8$ up-sampling of the LR image using bicubic interpolation. The rightmost column is the result of $\times 8$ up-sampling using our method.  
    \label{fig:results}}
\end{figure*}

As there is no effective and standard metric for perceptual quality, some representative qualitative results are presented in Fig.~\ref{fig:results2} and \ref{fig:results}. In Fig.~\ref{fig:results2}, we zoom in on smaller patches in the images, and in Fig.~\ref{fig:results} we show the full images. 

It can be observed from Fig.~\ref{fig:results2}~and~\ref{fig:results}, that our proposed model effectively outperforms standard interpolation in both details and sharpness. Considering the HR image is down-sampled to a spatial size of $32\times 32$ -- to allow for low-bandwidth image transmission -- and in that process loses much of the high-level detail, the HR image is recovered with remarkable results. However, it may be observed that the GAN-based approach sometimes introduces unwanted artifacts. More specifically, these can be observed by carefully observing the ``headshots'' in Fig.~\ref{fig:results2}.

\section{Conclusion and future work}
In this paper, we have investigated the prospect of applying Single Image Super-Resolution for practical maritime domain-specific use-cases to address some current challenges for wireless underwater data transmission. The focus of this work was the perceptual quality of the super-resolved reconstructed underwater images obtained from trawl fisheries. We have presented our GAN-based network architecture, and show representative qualitative results from experiments on our dataset. Our method outperforms standard interpolation methods, and show satisfying reconstruction of even high-frequency detail. We do however also notice some reconstruction artifacts, which we believe is a combination of the adversarial loss produced by the discriminator, and the very diverse image settings in the dataset. For use-cases where camera setting, position, and light may be controllable, we expect better recovery and fewer artifacts. Future work may involve collecting new datasets with camera systems where sensor and light settings are controllable to ensure less varying images and conduct practical experiments with these methods on underwater acoustic channels.

\section*{Acknowledgment}
We thank Dr. Daniel Stepputis, Head of the Fisheries and Survey Technology working group at the Thünen Institute of Baltic Sea Fisheries, and his group, for supplying datasets from their research.

\bibliographystyle{bib/IEEEtran}
\bibliography{bib/IEEEabrv,bib/bibliography}

\end{document}